\documentclass[11pt,a4paper]{article}

\usepackage{amsmath,amssymb,bm}
\usepackage{color}
\usepackage{multirow}

\usepackage{graphicx}
\usepackage{hyperref}

\usepackage{tikz}

\DeclareMathAlphabet{\mathpzc}{OT1}{pzc}{m}{it}

\newcommand{\bR}{\mathbb{R}}
\newcommand{\bZ}{\mathbb{Z}}

\newcommand{\cH}{\mathcal{H}}

\newcommand{\cN}{\mathcal{N}}
\newcommand{\cO}{\mathcal{O}}

\newcommand{\cZ}{\mathcal{Z}}

\newcommand{\U}{\mathrm{U}}

\newcommand{\SL}{\mathrm{SL}}

\newcommand{\Tr}{\operatorname{Tr}}

\newcommand{\Str}{\operatorname{Str}}

\newcommand{\dif}{\mathrm{d}}
\newcommand{\nap}{\mathrm{e}}
\newcommand{\im}{\mathrm{i}}

\makeatletter

\@addtoreset{equation}{section}
\makeatother

\date{\today}

\begin{document}

\begin{titlepage}

\renewcommand{\thefootnote}{\fnsymbol{footnote}}

\begin{flushright}
 {\tt 
 IPHT-T15/020 \\
 RIKEN-MP-110
 }
\\
\end{flushright}

\vskip9em

\begin{center}
 {\Large {\bf 
 Linking loops in ABJM and refined theory
 }}

 \vskip5em

 \setcounter{footnote}{1}
 {\sc Taro Kimura}\footnote{E-mail address: 
 \href{mailto:taro.kimura@cea.fr}
 {\tt taro.kimura@cea.fr}}

 \vskip2em

{\it 
 Institut de Physique Th\'eorique,
 CEA Saclay, 91191 Gif-sur-Yvette, France
 \\ \vspace{.5em}
 Mathematical Physics Laboratory, RIKEN Nishina Center, 
 Saitama 351-0198, Japan 
}

 \vskip3em

\end{center}

 \vskip2em

\begin{abstract}
 We consider the link average of the half-BPS Wilson loop
 operators in $\cN=6$ superconformal Chern--Simons--matter theory, which
 is called ABJM theory.
 We show that this loop average is reduced to a (super)matrix
 integral by the localization method, 
 in a similar way 
 to the bosonic $\U(N)$ Chern--Simons theory.
 Using this matrix integral, 
 we compute the two- and three-link averages 
 with an operator formalism inspired by a three-dimensional topological
 field theory.
 We obtain a factorization of the link average, and
 the Verlinde formula in a sector of supergroup representations.
 We also propose a refined version of 
 ABJM theory, and compute some refined link averages. 
\end{abstract}

\end{titlepage}

\tableofcontents

\hrulefill
\setcounter{footnote}{0}


\section{Introduction}\label{sec:intro}

The Wilson loop operator plays an important role in
Chern--Simons theory, which is a three-dimensional topological field theory.
This is because its expectation value is a topological invariant,
which encodes the shape of a knot along which the loop operator is
defined~\cite{Witten:1988hf}.
In addition to a single knot invariant, the average of linked loops
also has a meaning in its relation to the
two-dimensional conformal field theory (CFT).
For example, 
the matrix element of the modular $S$-matrix is
given by the two-link average, which is called the Hopf link invariant.
Furthermore, the three-link average also has an interpretation as the
fusion coefficient in the corresponding two-dimensional CFT.
In this way, link averages play a key role in the connection between
the three-dimensional topological theory and the two-dimensional CFT.

Recently a new class of Chern--Simons theories has been proposed based
on motivations from string and M-theory.
The most important example is $\cN=6$ superconformal
Chern--Simons--matter theory, which is called ABJM
theory~\cite{Aharony:2008ug,Aharony:2008gk}.
Although the direct computation of the path integral is a
difficult problem in general, we can reduce such an infinite-dimensional
integral to a matrix integral in a class of
supersymmetric field theories.
This procedure is called the localization method~\cite{Pestun:2007rz},
and shows that $\U(N)_k \times
\U(M)_{-k}$ ABJM theory reduces to $\U(N|M)$ Chern--Simons
theory~\cite{Kapustin:2009kz,Drukker:2009hy,Marino:2009jd}.
In this supergroup $\U(N|M)$, the bosonic part encodes the gauge
symmetry, and the fermionic part characterizes the matter content of the
theory~\cite{Gaiotto:2008sd}.
By applying various methods developed in matrix models, one can
compute 
Wilson loop averages in addition
to the partition function itself, 
in agreement with the result from the AdS/CFT correspondence.

In this paper we consider link averages of the half-BPS Wilson
loop operator in ABJM theory, based on the finite-dimensional matrix
integral expression, obtained via the localization method.
We will show that some important properties of the $\U(N)$ Chern--Simons theory
is generalized to the $\U(N|M)$ theory, at least in specific situations.


The bosonic Chern--Simons theory has an interesting
generalization 
called the refined Chern--Simons theory~\cite{Aganagic:2011sg}. 
In this case, the Wilson loop expectation values provide refined knot
invariants, which are given by the Poincar\'e polynomial of the
corresponding knot homology.
The refined theory has no Lagrangian description, but
only a construction based on topological string theory.
Following the same argument in the $\U(N|M)$
theory, we will consider the refinement of ABJM theory, and compute several
link averages of the Wilson loop operator in the refined theory.
In this case, these loop operators are described by the $\U(N|M)$ Macdonald
polynomial.

\section{Localizing linking loops}\label{sec:localization}

Computing the partition function and observable is highly non-trivial
in general, because one has to deal with infinite-dimensional path
integral in quantum field theory.
On the other hand, in a class of quantum field theories involving
supersymmetry, this infinite-dimensional integral can be reduced to
a matrix integral, by applying the so-called
localization method. 

It was shown by \cite{Kapustin:2009kz} that, in $\cN=2$ supersymmetric
Chern--Simons theory on three-sphere~$S^3$, the supersymmetrized Wilson
loop operator~\cite{Gaiotto:2007qi} along a great circle 
preserves a supersymmetry, which is necessary for localizing the path
integral. 
Thus its expectation value can be written in a form of
matrix integral.
For example, for $\cN=2$ Chern--Simons theory with gauge group $G =
\U(N)$, involving no matter fields, one can compute the Wilson loop average
\begin{align}
 \Big< W_R \Big>_{S^3}
 & = 
 \frac{1}{\cZ_{\U(N)}} \frac{1}{N!}
 \int \prod_{i=1}^N \frac{\dif x_i}{2\pi} \, \nap^{-\frac{1}{2g_s} x_i^2} \,
 \prod_{i<j}^N \left( 2 \sinh \frac{x_i - x_j}{2} \right)^2 \,
 \operatorname{Tr}_R U(x)
  \, ,
\end{align}
where $U(x) = \operatorname{diag}(\nap^{x_1},\ldots,\nap^{x_N})$ is a
holonomy matrix.
Since it is an element of $\U(N)$, these variables should be seen as
pure imaginary, $x_i \in \im \bR$.
Even with this parametrization, this integral is still converging,
because, as specified later, the coupling constant $g_s$ is also pure
imaginary.
Therefore it makes sense as the Fresnel integral.
Then the corresponding partition function is given by
\begin{align}
 \cZ_{\U(N)} & =
 \frac{1}{N!}
 \int \prod_{i=1}^N \frac{\dif x_i}{2\pi} \, \nap^{-\frac{1}{2g_s} x_i^2} \,
 \prod_{i<j}^N \left( 2 \sinh \frac{x_i - x_j}{2} \right)^2 
 \, .
 \label{CS_pf}
\end{align}
This result is consistent with the bosonic Chern--Simons
theory~\cite{Marino:2002fk}, because, in this case without matter
contributions, we can integrate out the auxiliary fields in a trivial way.

As pointed out in~\cite{Kapustin:2009kz}, the choice of great circle
is not unique.
There is a family of circles on which the Wilson loops preserve the same
supersymmetry used to localize the path integral.
These circles are generated by a vector field built with the Killing
spinor $\epsilon^\dag \gamma_\mu \epsilon$, and form a Hopf fibration.
Therefore one can similarly compute an average of linking loops in this
fibration by the localization using the same supersymmetry,
\begin{align}
 \Big<
  W_{R_1} \cdots W_{R_k}
 \Big>_{S^3}
 & =
 \Big<
  \Tr_{R_1} U(x) \cdots \Tr_{R_k} U(x)
 \Big>_{\U(N)}
 \, .
 \label{link_av1}
\end{align}
This is also consistent with the link average in the bosonic
Chern--Simons theory.
We remark that a multi-loop average in a four-sphere $S^4$ can be
obtained in a similar way, as multiple insertion of holonomy matrices
into the matrix integral~\cite{Pestun:2007rz}.
However these loops are not linked anymore in four-dimensional space.

In order to consider the maximally supersymmetric Wilson loop in ABJM
theory, namely the half-BPS Wilson loop operator in $\cN=6$ theory, one
has to assign a $\U(N|M)$ superconnection, instead of the bosonic
$\U(N)$ connection, 
which just provides the $1/6$ BPS Wilson loop~\cite{Drukker:2009hy}.
An important observation here is that these two
kinds of loops belong to the same cohomology class under the
supercharge used in the computation for $\cN=2$ theory.
Thus the difference between them is exact with respect to a linear
combination of the supercharges.
This implies that the same localization method can be applied to the half-BPS
linking loops forming the Hopf fibration, 
\begin{align}
 \Big<
  W_{R_1} \cdots W_{R_k}
 \Big>_{S^3}
 & =
 \Big<
  \Str_{R_1} U(x;y) \cdots \Str_{R_k} U(x;y)
 \Big>_{\U(N|M)}
 \, ,
 \label{link_av2}
\end{align}
where the corresponding partition function is the so-called ABJ(M) matrix
model, which is seen as a supermatrix version of the Chern--Simons matrix
model,
\begin{align}
 \cZ_{\U(N|M)}
 & =
 \frac{1}{N!M!} \int 
 \prod_{i=1}^N \frac{\dif x_i}{2\pi} \, \nap^{-\frac{1}{2g_s} x_i^2}
 \prod_{j=1}^M \frac{\dif y_j}{2\pi} \, \nap^{\frac{1}{2g_s} y_i^2} \,
 \prod_{i=1}^N \prod_{j=1}^M \left( 2 \cosh \frac{x_i-y_j}{2} \right)^{-2}
 \nonumber \\
 & \hspace{12em} \times
 \prod_{i<j}^N \left( 2 \sinh \frac{x_i-x_j}{2} \right)^2
 \prod_{i<j}^M \left( 2 \sinh \frac{y_i-y_j}{2} \right)^2
 \, ,
 \label{ABJM_pf}
\end{align}
and the supersymmetrized holonomy matrix is given by
\begin{align}
 U(x;y) = 
 \left(
  \begin{array}{cc}
   U(x) & \\ & -U(y)
  \end{array}
 \right)
 \, .
\end{align}
The trace of this matrix yields a character of $\U(N|M)$ group in
representation $R$, which is expressed by the Schur polynomial with a
prescribed symmetry~\cite{Berele:1987yi},
\begin{align}
 \Str_R U(x;y) & = s_{\lambda(R)} (\nap^x;\nap^y) \, ,
 \label{SUSY_Schur}
\end{align}
where $\lambda(R)$ is the highest weight vector corresponding to the
representation $R$.
This character is obtained from the Schur polynomial for
$\U(N+M)$ group in the following way:
The Schur polynomial can be expressed as a linear combination of 
power-sum polynomials $p_\mu(x,y)=\prod_{j}p_{\mu_j}(x,y)$,
corresponding to a trace in the fundamental representation $p_n(x,y) =
\Tr U(x,y)^n$, which is known as the Frobenius formula,
\begin{align}
 s_\lambda(x,y) 
 & = 
 \sum_{\mu} \frac{1}{z_\mu} \chi_\lambda(C_\mu) \, p_{\mu}(x,y)
 \, ,
\end{align}
where $\chi_\lambda$ and $C_\mu$ are the character and the conjugacy
class for the symmetric group $\mathfrak{S}_{N+M}$, and the coefficient
$z_\mu$ is given by $z_\mu = \prod_j \mu_j! \, j^{\mu_j}$.
Replacing the ordinary trace with the supertrace $p_n(x,y) \to p_n(x;y)
= \Str U(x;y)^n$ in this expression, one obtains (\ref{SUSY_Schur}).
See Appendix~\ref{sec:Schur} for various properties of this Schur
polynomial, which will be used in the following Sections.

\section{Operator formalism}\label{sec:op_formalism}

As shown in Section~\ref{sec:localization}, the link average in ABJM
theory can be discussed in a similar way to the bosonic
Chern--Simons theory, which is a topological field theory in three
dimensions.
In general, the path integral of topological field theory on a
three-manifold $M$ can be described using a state in the Hilbert space
$\cH_{\Sigma}$ associated with the boundary $\partial M = \Sigma$, which is
obtained through canonical quantization on $\Sigma\times\bR$.
Once a state in this Hilbert space is given, $|M\rangle \in \cH_\Sigma$, its
dual is obtained by inverting the orientation of the boundary, $\langle
M| \in \cH_\Sigma^*$.
Then the inner product of these states gives the partition function of
the three-dimensional theory on $M$.
In this sense it is convenient to consider the operator formalism based on
this boundary theory in order to compute observables in topological field
theory on the three-manifold $M$.
We will formally apply this construction to ABJM theory, which is
motivated by the bosonic Chern--Simons theory.

\subsection*{Loop insertion and modular transformation}

Let us first review the operator formalism in the bosonic
Chern--Simons theory~\cite{Witten:1988hf}.
For our purpose, we choose the boundary as a two-dimensional torus
$\Sigma=T^2$, and take a solid torus to obtain the Hilbert space
$\cH_{T^2}$.
We consider a state given by inserting a Wilson line in representation
$R_i$ along the non-contractible cycle of the solid torus,
\begin{align}
 | R_i \rangle  
 = \cO_{R_i} | 0 \rangle
 \in \cH_{T^2}
 \, ,
\end{align}
where the operator $\cO_{R_i}$ is a creation operator of the Wilson loop
in representation $R_i$.
Then taking an inner product between this and its dual, we obtain the
partition function with two Wilson lines in $S^2 \times S^1$,
\begin{align}
 \left< R_i | R_j \right>
 & =
 \cZ(S^2 \times S^1;\bar{R}_i,R_j)
 \, .
\end{align}
This defines a Hermitian metric
\begin{align}
 g_{\bar{i}j} = \left< R_i | R_j \right>
 \label{H_metric}
\end{align}
on the Hilbert space, because its
conjugation is simply given by $g_{\bar{i}j}^* = g_{\bar{j}i}$.

In addition to the creation operator $\cO_{R}$ of the Wilson line, we
have another set of operators, corresponding to the mapping
class group of the torus, which is given by $\SL(2,\bZ)$ group.
Let us introduce $S$ and $T$ matrices, describing the modular
transformation of the torus, with the $\SL(2,\bZ)$ relation,
\begin{align}
 S^4 = 1  \, , \quad
 (ST)^3 = S^2 \, .
\end{align}
These allow us to compute the path integral on a three-sphere $S^3$.
This is because the three-sphere $S^3$ is obtained from $S^2 \times S^1$
by applying the modular transformation $S$, which converts the modulus of the
torus, $\tau \to -1/\tau$.
From this point of view, the matrix element of the $S$-matrix computes
the linking knots, namely the Hopf link average in the
three-sphere~$S^3$,
\begin{align}
 S_{\bar{i}j}
 & =
 \langle R_i | S | R_j \rangle
 =
 \cZ(S^3;\bar{R}_i,R_j)
 \, ,
 \label{S-mat}
\end{align}
and, by normalizing it with the partition function, we obtain the Hopf
link invariant
\begin{align}
 \frac{S_{\bar{i}j}}{S_{00}}
 & =
 \frac{\cZ(S^3;\bar{R}_i,R_j)}{\cZ(S^3)}
 \, .
\end{align}
We will compute this average in Section~\ref{sec:2pt}.
We remark that, to obtain the three-sphere $S^3$, we can apply more
general choice of the transformation, $T^n S T^m$ with arbitrary
integers $n$ and $m$.
This $T$-transformation gives rise to a framing factor in the knot
average.

\subsection*{Wavefunction}

Based on the formalism discussed above, let us then consider a
wavefunction corresponding to a state in the Hilbert space, which
allows us to compute the path integral more explicitly.

For the $\U(N)$ Chern--Simons theory, the wavefunction for the situation
without any Wilson lines, corresponding to the vacuum state, is given by the
partition function of Chern--Simons theory on a solid
torus~\cite{Aganagic:2002wv}.
Let us denote it by
\begin{align}
 \langle\,x \, | \, 0 \, \rangle
 & =
 \frac{1}{\sqrt{N!}}
 \prod_{i<j}^N \left( 2 \sinh \frac{x_i-x_j}{2} \right)
 \, .
\end{align}
The factor is determined to be consistent with normalization of the
metric.
Because the expression on the right-hand side can be written as the
Vandermonde determinant, up to some trivial factors, this wavefunction
is seen as the Slater determinant of the $N$-particle system.

If we insert the Wilson line in representation $R$ into the solid torus,
the corresponding wavefunction is given by
\begin{align}
 \langle\,x \, | \, R\,\rangle
 & =
 \frac{1}{\sqrt{N!}}
 \prod_{i<j}^N \left( 2 \sinh \frac{x_i-x_j}{2} \right) \, 
 s_R (\nap^x)
 \, .
\end{align}
The Schur polynomial $s_R(\nap^x)$ added here shows the
trace of the holonomy matrix in representation $R$,
\begin{align}
 \Tr_R U(x) = s_R(\nap^x) \, .
\end{align}
Now the highest weight vector for the representation $R$ is also denoted
by $R$, as far as there is no confusion.
As pointed out in Section~\ref{sec:localization}, these
variables shall behave as pure imaginary $x_i \in \im \bR$.
Thus the conjugate of this wavefunction is given by
\begin{align}
 \langle \, R \, | \, x \, \rangle
 & =
 (-1)^{N(N-1)/2}
 \frac{1}{\sqrt{N!}}
 \prod_{i<j}^N \left( 2 \sinh \frac{x_i-x_j}{2} \right) \, 
 s_R (\nap^{-x})
 \, .
\end{align}
Then we compute the inner product of these states,
\begin{align}
 \langle R_i | R_j \rangle
 & =
 \frac{1}{N!} \int \prod_{i=1}^N \frac{\dif x_i}{2 \pi} \,
 \prod_{i<j}^N \left( 2 \sinh \frac{x_i-x_j}{2} \right)^2 \,
 s_{R_i} (\nap^{-x}) s_{R_j} (\nap^x)
 \, .
\end{align}
Applying the determinantal formula for the Schur polynomial
\begin{align}
 s_\lambda(x) & =
 \frac{1}{\Delta(x)}
 \det_{1 \le i, j \le N} x_i^{\lambda_j+N-j}
 \quad \text{with} \quad
 \Delta(x) = \prod_{i<j}^N (x_i - x_j)
 \, ,
 \label{Schur_det2}
\end{align}
the inner product is given by
\begin{align}
 \langle R_i | R_j \rangle 
 & =
 \det_{1 \le k, l \le N}
 \left(
  \int_{\im \bR} 
   \frac{\dif x}{2\pi} \, \nap^{x (R_{j,k} - R_{i,l} - k + l)}
 \right)
 \, ,
\end{align}
which yields
\begin{align}
 g_{\bar{i}j} = \delta^i_{~j} \, .
\end{align}
Here $R_{j,k}$ stands for the $k$-th component of the highest weight
vector corresponding to the representation $R_j$.
This is just the orthonormal property of the Schur polynomial with respect
to the weight function $\Delta(\nap^x)^2$~\cite{Macdonald:1997}.

\subsection*{$\U(N|M)$ theory}

Let us then discuss the $\U(N|M)$ Chern--Simons theory in a similar way.
Actually, in the sense of ABJM theory, the meaning of the operator
formalism is not yet clear, because this ABJM theory is not just a
topological theory.
However, there are several results encouraging us to apply it to ABJM theory.
For example, the ABJM matrix model is equivalent to Chern--Simons theory
on the lens space $S^3/\bZ_2$ through the analytic
continuation~\cite{Marino:2009jd}, and at least for the latter theory,
we can apply the same operator formalism because the lens space Chern--Simons
theory is a topological theory defined on a Seifert-type three-manifold.
Another remark is that the ABJM model can be seen as a many-body system
of fermionic particles~\cite{Marino:2011eh}.
From this point of view, it is natural to consider the wavefunction of
ABJM theory in a similar way to the bosonic $\U(N)$ theory shown above.

Now, by analogy with the $\U(N)$ Chern--Simons theory, we introduce a
wavefunction for the $\U(N|M)$ theory, corresponding to the vacuum state
\begin{align}
 \langle \, x;y \, | \, 0 \, \rangle
 & =
 \frac{1}{\sqrt{N!M!}}
 \prod_{i<j}^N \left( 2 \sinh \frac{x_i-x_j}{2} \right)
 \prod_{i<j}^M \left( 2 \sinh \frac{y_i-y_j}{2} \right)
 \prod_{i=1}^N \prod_{j=1}^M \left( 2 \cosh \frac{x_i-y_j}{2} \right)^{-1} 
 \, .
\end{align}
As well as the bosonic Chern--Simons theory, this wavefunction
can be interpreted as the Slater determinant based on the formula,
\begin{align}
 &
 \prod_{i<j}^N \left( 2 \sinh \frac{x_i-x_j}{2} \right)
 \prod_{i<j}^M \left( 2 \sinh \frac{y_i-y_j}{2} \right)
 \prod_{i=1}\prod_{j=1}^M \left( 2 \cosh \frac{x_i-y_j}{2} \right)^{-1}
 \nonumber \\
 & =
 \prod_{i=1}^N \nap^{\frac{-N+M+1}{2}x_i}
 \prod_{j=1}^M \nap^{\frac{N-M+1}{2}y_j}
 \det 
 \left(
  \begin{array}{c}
   \nap^{x_i (k-1)} \\ \left( \nap^{x_i} + \nap^{y_j} \right)^{-1}
  \end{array}
 \right)
 \quad \text{with} \quad
 \begin{cases}
  i = 1,\ldots,N \\
  j = 1,\ldots,M \\
  k = 1,\ldots,N-M
 \end{cases}
 \, ,
\end{align}
where we assume $N \ge M$.
Similarly a wavefunction for a state with the Wilson line in
representation $R$ is given by
\begin{align}
 \langle \, x; y \, | \, R\, \rangle
 & =
 \frac{1}{\sqrt{N!M!}}
 \prod_{i<j}^N \left( 2 \sinh \frac{x_i-x_j}{2} \right)
 \prod_{i<j}^M \left( 2 \sinh \frac{y_i-y_j}{2} \right)
 \prod_{i=1}^N \prod_{j=1}^M \left( 2 \cosh \frac{x_i-y_j}{2} \right)^{-1} 
 s_{R}(\nap^x;\nap^y)
 \, , \\
 \langle \, R \, | \, x; y \, \rangle
 & =
 \frac{1}{\sqrt{N!M!}}
 \prod_{i<j}^N \left( 2 \sinh \frac{x_i-x_j}{2} \right)
 \prod_{i<j}^M \left( 2 \sinh \frac{y_i-y_j}{2} \right)
 \prod_{i=1}^N \prod_{j=1}^M \left( 2 \cosh \frac{x_i-y_j}{2} \right)^{-1} 
 s_{R}(\nap^{-x};\nap^{-y})
 \, .
\end{align}
Although it is difficult to compute the inner product for generic
representations, an interesting simplification occurs in a special case:
If the highest weight vector $\lambda$ satisfies $\lambda_N \ge
M$, the $\U(N|M)$ Schur polynomial is factorized into the $\U(N)$ and
$\U(M)$ Schur polynomials.
We call such a representation {\em a maximal representation} in this paper.%
\footnote{
The $\U(N|M)$ Schur polynomial becomes identically zero in the
case with $\lambda_{N+1} \ge M+1$.
In this sense, the representation with $\lambda_N \ge M$ is
interpreted as maximal, and this is the reason why we call the situation
with $\lambda_N \ge M$ a maximal representation.
See Appendix~\ref{sec:Schur} and also~\cite{Berele:1987yi,Moens:2003JAC}.
}
We focus on this situation in the following.

Let us compute the inner product of the states defined above.
When both of the representations belong to the maximal class, we can apply
the formula \eqref{Schur_fact02}.
In such a case it is given by
\begin{align}
 \langle R_i | R_j \rangle
 & =
 \det_{1 \le k, l \le N}
 \left( 
  \int \frac{\dif x}{2\pi} \, \nap^{x(R_{j,k}^{(1)}-R_{i,l}^{(1)}-k+l)}
 \right)
 \det_{1 \le k, l \le M}
 \left( 
  \int \frac{\dif y}{2\pi} \, \nap^{y(R_{j,k}^{(2)}-R_{i,l}^{(2)}-k+l)}
 \right)
 \nonumber \\
 & =
 \delta_{R_i^{(1)} R_j^{(1)}} \delta_{R_i^{(2)} R_j^{(2)}}
 \, ,
 \label{inner_prod}
\end{align}
where the representations $R_i^{(1)}$ and $R_i^{(2)}$ are obtained from
the original one so that
$R_{i,k}^{(1)}=R_{i,k}-M$ for $k=1,\ldots,N$ and
$R_{i,k}^{(2)}=R_{i,k}^t-N$ for $k=1,\ldots,M$.
See Figure~\ref{fig:partition}.
This shows the orthogonality relation for $\U(N|M)$ Schur polynomial
with respect to the corresponding weight function, and thus the metric
\eqref{H_metric} is given by
\begin{align}
 g_{\bar{i}j} & = \delta^i_{~j}
 \, .
\end{align}
We have derived this result only for the maximal
representation.
However it is not obvious whether this orthogonality holds for
generic representations.
We will give a relating comment in Section~\ref{sec:discussion} in the
relation to the two-dimensional CFT with internal supersymmetry.

\subsection{Two-point function: Hopf link}\label{sec:2pt}

As mentioned before, 
the $S$-matrix plays a
key role in computing the path integral for the three-sphere $S^3$.
Actually its matrix element provides the Hopf link invariant in $S^3$,
and also involves the unknot invariant as a special case.
In this Section we consider the $S$-matrix element in an explicit
way, and then compute the Hopf link average as the two-point function in
the ABJM matrix model.

\subsection*{$\U(N)$ theory}

For the $\U(N)$ theory, the wavefunction corresponding to the state obtained
by the modular transformation $TST$ is given by~\cite{Aganagic:2002wv}
\begin{align}
 \langle \, x \, | \, TST \, | \, R \, \rangle
 & =
 \frac{1}{\sqrt{N!}} 
 \prod_{i=1}^N \nap^{-\frac{1}{2g_s}x_i^2} \,
 \prod_{i<j}^N \left( 2 \sinh \frac{x_i-x_j}{2} \right) \,
 s_R(\nap^x)
 \, ,
\end{align}
where $g_s$ is the coupling constant defined with the level and the rank of
Chern--Simons theory,
\begin{align}
 g_s & = \frac{2\pi \im}{k+N} \, .
\end{align}
As a result, we can compute the matrix element of $TST$ as a two-point
function of the Schur polynomial in the Chern--Simons matrix model,
\begin{align} 
 (TST)_{\bar{i}j} 
 & =  \langle \, R_i \, | \, TST \, | \, R_j \, \rangle
 \nonumber \\
 & =
 \frac{1}{N!}
 \int \prod_{i=1}^N \frac{\dif x_i}{2\pi} \,
 \nap^{-\frac{1}{2g_s} x_i^2} \,
 \prod_{i<j}^N \left( 2 \sinh \frac{x_i-x_j}{2} \right)^2 \,
 s_{R_i}(\nap^{-x}) s_{R_j}(\nap^x) 
 \, .
\end{align}
This is consistent with the expression \eqref{link_av1}, up to the
orientation of the loop, because the operator in the representation
$R_i$ is now in the opposite direction.
This two-point function can be easily computed using the explicit
formula of the Schur polynomial \eqref{Schur_det2},
\begin{align}
 \det_{1 \le m, n \le N}
 \left(
  \int \frac{\dif x}{2\pi} \,
  \nap^{-\frac{1}{2g_s} x^2 + x(R_{j,m} - R_{i,n} - m + n)}
 \right)
 & = 
 \left( \frac{g_s}{2\pi} \right)^{N/2}
 \det_{1 \le m, n \le N} 
 q^{-\frac{1}{2}(R_{j,m} - R_{i,n} - m + n)^2}
 \, .
\end{align}
We have defined the parameter $q$ by $q=\nap^{-g_s}$.
If we modify the variables $x_i \to \im x_i$, it should be replaced with
$q \to \nap^{g_s}$.

Rewriting the expression in terms of the Schur polynomial again, it yields
\begin{align}
 (TST)_{\bar{i}j}
 & =
 c_N \,
 q^{-\frac{1}{2}(C_2(R_i)+C_2(R_j))}
 \Delta(q^\rho) \, s_{R_i}(q^{R_j+\rho}) \, s_{R_j}(q^{\rho})
 \, ,
\end{align}
where the constant $c_N$ is given by $c_N = (g_s/(2\pi))^{N/2}
q^{\frac{1}{12}N(N-1)(2N-1)}$, $\rho$ is the Weyl vector, $\rho_i = - i
+ 1/2$, and $C_2(R)$ is the second Casimir operator,
\begin{align}
 C_2(R) & =
 \sum_{k=1}^N
 \left(
  \left( R_k - k + \frac{1}{2} \right)^2 -
  \left( - k + \frac{1}{2} \right)^2
 \right)
 \, .
\end{align}
Actually the factor of this Casimir operator is interpreted as the
framing factor, which is given by the conformal weight of the primary
field, 
and this contribution just stands for the action of the modular
$T$-matrix,
\begin{align}
 T_{\bar{i}j} & = \delta^i_{~j} \, q^{-\frac{1}{2}C_2(R_i)} \, .
\end{align}
Thus we obtain the Hopf link invariant by the normalized $S$-matrix,
\begin{align}
 \frac{S_{\bar{i}j}}{S_{00}}
 & = 
 s_{R_i} (q^{R_j+\rho}) s_{R_j} (q^\rho)
 \, .
 \label{S-mat_Schur1}
\end{align}
Although this expression is not symmetric superficially, we can show
that it is symmetric under the exchange $i \leftrightarrow j$.

\subsection*{$\U(N|M)$ theory}

Let us then apply the same approach to the $\U(N|M)$ theory.
In this case, although there is no rigorous foundation of this argument,
we propose the following form of the wavefunction based on
consistency and analogy with the $\U(N)$ theory,
\begin{align}
 \langle \, x; y \, | \, TST \, | \, R \, \rangle
 & =
 \frac{1}{\sqrt{N!M!}} 
 \prod_{i=1}^N \nap^{-\frac{1}{2g_s} x_i^2} \,
 \prod_{j=1}^M \nap^{\frac{1}{2g_s} y_i^2} \,
 \prod_{i=1}^N \prod_{j=1}^M
 \left( 2 \cosh \frac{x_i - y_j}{2} \right)^{-1}
 \nonumber \\
 & \hspace{3em} \times
 \prod_{i<j}^N \left( 2 \sinh \frac{x_i-x_j}{2} \right) \,
 \prod_{i<j}^M \left( 2 \sinh \frac{y_i-y_j}{2} \right) \,
 s_R(\nap^x;\nap^y)
 \, .
\end{align}
Actually this expression is obtained from the Chern--Simons theory on
the lens space $S^3/\bZ_2$ through the analytic continuation.
Now the coupling constant depends only on the level of $\U(N)_k \times
\U(M)_{-k}$ ABJ(M) theory,
\begin{align}
 g_s & = \frac{2\pi \im}{k} \, .
\end{align}
Using this expression, we obtain the corresponding matrix element,
which is consistent with the Hopf link average \eqref{link_av2} in this
theory
\begin{align}
 \langle \, R_i \, | \, TST \, | \, R_j \, \rangle
 & =
 \frac{1}{N!M!} \int 
 \prod_{i=1}^N \frac{\dif x_i}{2\pi} \, \nap^{-\frac{1}{2g_s} x_i^2}
 \prod_{j=1}^M \frac{\dif x_j}{2\pi} \, \nap^{\frac{1}{2g_s} y_j^2}
 \prod_{i=1}^N \prod_{j=1}^M
 \left( 2 \cosh \frac{x_i - y_j}{2} \right)^{-2}
 \nonumber \\
 & \quad \times
 \prod_{i<j}^N \left( 2 \sinh \frac{x_i - x_j}{2} \right)^2
 \prod_{i<j}^M \left( 2 \sinh \frac{y_i - y_j}{2} \right)^2 \,
 s_{R_i}(\nap^{-x};\nap^{-y}) \, s_{R_j}(\nap^x;\nap^y)
 \, .
 \label{TST_ME2}
\end{align}
We remark that the orientation of the loop operator in the
representation $R_i$ is flipped again.
Let us denote this matrix element by $(TST)_{\bar{i}j}^{\U(N|M)}$, to
distinguish it from that in the bosonic $\U(N)$ theory.
We can immediately check that this reproduces the $\U(N|M)$ matrix model
partition function \eqref{ABJM_pf} by taking the trivial representation, 
\begin{align}
 (TST)_{00}^{\U(N|M)} = \cZ_{\U(N|M)}
 \, .
\end{align}

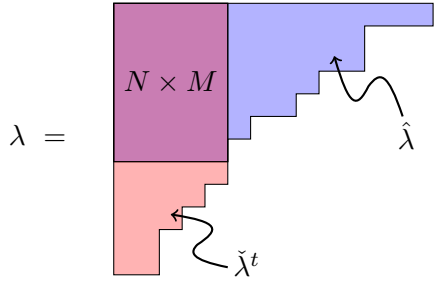
\begin{figure}[t]
 \begin{center}
  \begin{tikzpicture}

   \filldraw [fill=blue,fill opacity=.3,draw=black]
   (1,2.1) -- (2.5,2.1) -- (5.2,2.1) -- (5.2,1.8) -- (4.3,1.8) -- 
   (4.3,1.2) -- (3.7,1.2) -- (3.7,0.9) -- (3.4,0.9) -- (3.4,0.6) --
   (2.8,0.6) -- (2.8,0.3) -- (2.5,0.3) -- (2.5,0) -- (1,0) -- cycle;

   \filldraw [draw=none,fill=red,fill opacity=.3,draw=black]
   (1,2.1) -- (1,0) -- (1,-1.5) -- (1.6,-1.5) -- (1.6,-0.9) -- (1.9,-0.9)
   -- (1.9,-0.6) -- (2.2,-0.6) -- (2.2,-0.3) -- (2.5,-0.3) 
   -- (2.5,0) -- (2.5,2.1) -- cycle;

   \draw 
   (1,0) -- (1,2.1) -- (2.5,2.1) -- (2.5,0) -- cycle ;

   \draw [<-,thick] (3.9,1.4) .. controls (4.1,0.7) and (4.2,0.7)
   .. (4.35,1) .. controls (4.5,1.2) and (4.6,1.2) .. (4.8,0.6);

   \draw [<-,thick] (1.75,-0.7) .. controls (2.4,-0.8) and (2.4,-0.9)
   .. (2.1,-1.1) .. controls (2,-1.2) and (2,-1.3) .. (2.5,-1.4);

   \path (0,0.35) node {$\lambda \ =$};
   \path (1.75,1.05) node {$N \times M$};
   \path (4.85,0.35) node {$\hat{\lambda}$};
   \path (2.75,-1.35) node {$\check{\lambda}^t$};

  \end{tikzpicture}
 \end{center}
 \caption{The partition $\lambda = (14,11,11,9,8,6,5,5,4,3,2,2)$ 
 satisfying the maximal condition $\lambda_N \ge M$ for $N=7$ and
 $M=5$, which includes $\hat{\lambda}=(14,11,11,9,8,6,5)$ (blue) and
 $\check{\lambda}=(12,12,10,9,8)$ (red).
 See also Figure~\ref{fig:partition}.
 }
 \label{fig:partition2}
\end{figure}

In general, it is difficult to compute this matrix element for
arbitrary representation.
However, a similar simplification occurs if we take the maximal
representation, as in the case of the inner product discussed before.
When one of the representations is trivial, and the other is maximal, it
gives the $\U(N|M)$ unknot average~\cite{Eynard:2014rba},
\begin{align}
 (TST)_{0R} 
 & =
 \frac{\im^{\frac{N-M}{2}}}{k^{\frac{N+M}{2}}} \,
 q^{-\frac{1}{6}(N-M)^3+\frac{1}{24}(N-M)}
 \prod_{i=1}^N q^{-\frac{1}{2}(\xi_i^2+\xi_i)}
 \prod_{j=1}^M q^{\frac{1}{2}(\eta_j^2+\eta_j)} \,
 \nonumber \\
 & \hspace{3em} \times
 \prod_{i=1}^N \prod_{j=1}^M
 \left( q^{-\xi_i} + q^{\eta_j} \right)^{-1}
 \prod_{i<j}^N \left( q^{-\xi_i} - q^{-\xi_j} \right)
 \prod_{i<j}^M \Big( q^{\eta_i} - q^{\eta_j} \Big)
 \, ,
\end{align}
where we have defined
\begin{align}
 \xi_i & = R_i - i + \frac{1}{2} = \hat{R}_i + \rho_i 
 \qquad \text{for} \quad i = 1,\ldots, N \, , \\
 \eta_j & = R_j^t - j + \frac{1}{2} = \check{R}_j + \rho_j
 \qquad \text{for} \quad j = 1,\ldots, M \, ,
\end{align}
with $\hat{R}=(R_i,\ldots,R_N)$ and
$\check{R}=(R_1^t,\ldots,R_M^t)$ as shown in Figure~\ref{fig:partition2}.
This average can be also written as a determinant, and reproduces the
$\U(N)$ unknot average by taking $M=0$.
For the latter convenience, let us assume $N=M$ and rewrite this
expression in terms of the Schur polynomial,
\begin{align}
 (TST)_{0R} 
 & =
 \frac{1}{k^N} \,
 q^{-\frac{1}{2}C_2(\hat{R})+\frac{1}{2}C_2(\check{R})}
 \Delta_N(q^{\rho}) \, \Delta_M(q^{-\rho}) \,
 s_{\hat{R}}(q^{\rho}) \, s_{\check{R}}(q^{-\rho})
 \prod_{i=1}^N \prod_{j=1}^M
 \left( q^{-\hat{R}_i-\rho_i} + q^{\check{R}_j+\rho_j} \right)^{-1} 
 \, .
 \label{TST_0R}
\end{align}

When the both of representations are maximal, we can again apply the
formula \eqref{Schur_fact02} to compute the matrix element
\eqref{TST_ME2}.
In this case we obtain an expression which is completely factorized
into $\U(N)$ and $\U(M)$ sectors,
\begin{align}
 (TST)_{\bar{i}j}
 & =
 \frac{\im^{\frac{N-M}{2}}}{k^{\frac{N+M}{2}}} \,
 \det_{1 \le k, l \le N}
 \left(
  q^{-\frac{1}{2}(\hat{R}_{j,k}-\hat{R}_{i,l}-k+l)^2}
 \right)
 \det_{1 \le k, l \le M}
 \left(
  q^{\frac{1}{2}(\check{R}_{j,k}-\check{R}_{i,l}-k+l)^2}
 \right)
 \, .
\end{align}
It is also expressed in terms of Schur polynomials,
\begin{align}
 (TST)_{\bar{i}j}
 & =
 c_{N,M} \,
 q^{-\frac{1}{2}(C_2(\hat{R}_i)+C_2(\hat{R}_j)
    +\frac{1}{2}(C_2(\check{R}_i)+C_2(\check{R}_j))}
 \nonumber \\
 & \qquad \times
 \Delta_N(q^\rho) \, \Delta_M(q^{-\rho}) \,
 s_{\hat{R}_i}(q^{\hat{R}_j+\rho}) \, s_{\hat{R}_j}(q^{\rho}) \,
 s_{\check{R}_i}(q^{-\check{R}_j-\rho}) \, s_{\check{R}_j}(q^{-\rho}) 
 \, ,
 \label{TST_sSchur}
\end{align}
with the constant factor
\begin{align}
 c_{N,M} = 
 \frac{\im^{\frac{N-M}{2}}}{k^{\frac{N+M}{2}}} \,
 q^{\frac{1}{12}N(N-1)(2N-1)-\frac{1}{12}M(M-1)(2M-1)}
 \, .
\end{align}
From the expressions of \eqref{TST_0R} and \eqref{TST_sSchur}, it is
natural to read off the matrix element of the $T$-matrix for the maximal
representation, which gives the framing factor in the loop average,
\begin{align}
 T_{\bar{i}j}^{\U(N|M)}(q) & = \delta^i_{~j} \, 
 q^{-\frac{1}{2} C_2(\hat{R}_i) + \frac{1}{2} C_2(\check{R}_i)}
 \, .
\end{align}
Thus we obtain the factorized $S$-matrix element,
\begin{align}
 S_{\bar{i}j}^{\U(N|M)}(q)
 & =
 S_{\bar{\hat{i}}\hat{j}}^{\U(N)}(q) \times
 S_{\bar{\check{i}}\check{j}}^{\U(M)}(q^{-1})
 \, .
\end{align}
This interesting property can be shown only for the maximal case at this moment.
It is expected that such a factorization is not found explicitly for
the non-maximal situation, because it is difficult to split the
$\U(N|M)$ representation into $\U(N)$ and $\U(M)$ sectors in general,
and probably related to the chiral non-factorizability of the
two-dimensional CFT with internal supersymmetry.
See also discussion in Section~\ref{sec:discussion}.

\subsection{Three-point function: Verlinde formula}\label{sec:3pt}

In addition to the two-point function, which gives the Hopf link
average, the three-point function also plays an important role in
the two-dimensional CFT, and the three-dimensional topological
field theory.
In CFT, the product of operators can be expanded by a set of operators
in general,
\begin{align}
 \cO_{R_i} \cO_{R_j}
 & = \sum_{k} N_{ij}^{~~k} \, \cO_{R_k}
 \, ,
 \label{OPE}
\end{align}
and thus the fusion coefficient $N_{ij}^{~~k}$ in this expansion plays a
role of the structure constant.
The statement of Verlinde's conjecture is that the $S$-matrix
diagonalizes this fusion coefficient, which is
equivalent to the relation for the $S$-matirx~\cite{Verlinde:1988sn},
\begin{align}
 \frac{S_{\bar{k}i} S_{\bar{k}j}}{S_{\bar{k}0}}
 & =
 \sum_{\ell} N_{ij}^{~~\ell} S_{\bar{k}\ell}
 \, .
 \label{V_formula}
\end{align}
Actually this coefficient is simply understood in terms of the operator
formalism,
\begin{align}
 N_{ijk} & =
 \langle 0 | R_i \, R_j \, R_k \rangle \, ,
\end{align}
which corresponds to the three-point function in Chern--Simons theory.
This is very general result for CFT, and topological
field theory associated with this fusion rule.

In particular, for the $\U(N)$ Chern--Simons theory, the fusion coefficient
\eqref{OPE} coincides with the Littlewood--Richardson coefficient, which
appears in the product of Schur polynomials,
\begin{align}
 s_{R_i} (x) \, s_{R_j} (x)
 & = 
 \sum_{k} N_{ij}^{~~k} \, s_{R_k} (x)
 \, .
 \label{LR01}
\end{align}
Now the degree of representation is conserved on the both-hand sides,
$|R_i|+|R_j|=|R_k|$.
This follows the fact that the action of the operator $\cO_{R_i}$ is just
given by multiplication of the Schur polynomial in representation $R_i$.
From this point of view, we can show the Verlinde formula
\eqref{V_formula} using the explicit form of the $S$-matrix element
\eqref{S-mat_Schur1}.
Multiplying the equation \eqref{V_formula} by $(S_{\bar{k}0})^{-1}$, the
left-hand side is given by
\begin{align}
 \frac{S_{\bar{k}i}}{S_{\bar{k}0}}
 \frac{S_{\bar{k}j}}{S_{\bar{k}0}}
 & =
 s_{R_i}(q^{R_k+\rho}) \, s_{R_j}(q^{R_k+\rho})
 =
 \sum_\ell N_{ij}^{~~\ell} \, s_{R_\ell}(q^{R_k+\rho}) \, ,
 \label{V_formula2}
\end{align}
which coincides with the right-hand side of the formula \eqref{V_formula}.

Let us then apply this argument to the $\U(N|M)$ theory.
Again we consider the maximal representations.
In this case it can be shown that the product of $\U(N|M)$ Schur polynomials
is expanded only with the maximal $\U(N|M)$ Schur polynomials again.
Applying the formula
\begin{align}
 \prod_{i=1}^N \prod_{j=1}^M (x_i + y_j)
  = \sum_{\lambda \subset M^N} s_\lambda(x) \, s_{\tilde{\lambda}^t} (y)
  \label{Schur_formula_rect}
\end{align}
where $\lambda \subset M^N$ implies $\lambda_1 \le M$ and $\lambda_1^t
\le N$ with $\tilde{\lambda}^t=(N-\lambda_M^t, \ldots, N -
\lambda_1^t)$~\cite{Macdonald:1997}, the product is given by
\begin{align}
 s_{R_i}(x;y) \, s_{R_j}(x;y)
 & =
 \sum_{\ell,m,n} 
 N_{i^{(1)}j^{(1)}}^{~~~~~\ell^{(1)}} \, N_{i^{(2)}j^{(2)}}^{~~~~~\ell^{(2)}} \,
 N_{\ell^{(1)}m}^{~~~~n^{(1)}} \, N_{\ell^{(2)}\tilde{m}^t}^{~~~~n^{(2)}} \,
 s_{R_n^{(1)}}(x) \, s_{R_n^{(2)}}(y) \,
 \prod_{i=1}^N \prod_{j=1}^M
 \left( x_i + y_j \right)
 \nonumber \\
 & =
 \sum_{n} \mathpzc{N}_{~ij}^{~~n} \, s_{R_n}(x;y)
 \, ,
 \label{LR02}
\end{align}
where the fusion coefficient is defined
\begin{align}
 \mathpzc{N}_{~ij}^{~~n}
 & =
 \sum_{\ell,m} 
 N_{i^{(1)}j^{(1)}}^{~~~~~\ell^{(1)}} \, N_{i^{(2)}j^{(2)}}^{~~~~~\ell^{(2)}} \,
 N_{\ell^{(1)}m}^{~~~~n^{(1)}} \, N_{\ell^{(2)}\tilde{m}^t}^{~~~~n^{(2)}} 
 \, .
\end{align}
Again we have a conservation law of the degree of representations,
$|R_i|+|R_j|=|R_k|$.
We remark that, since the representation is now restricted to the maximal
ones, the summation over the representation means 
$\sum_n = \sum_{n^{(1)},n^{(2)}}$ and so on.
The fusion coefficient \eqref{LR02} is the $\U(N|M)$ version of the
Littlewood--Richardson coefficient, and this shows that the maximal
representations form a closed subsector in the whole space of
$\U(N|M)$ representations.

Once the fusion coefficient is given in the $\U(N|M)$ theory, we can
similarly discuss the Verlinde formula for the modular $S$-matrix
\eqref{V_formula}.
Assuming $N=M$ for simplicity, the ratio of the $S$-matrix is obtained
from the expressions \eqref{TST_0R} and \eqref{TST_sSchur}, 
\begin{align}
 \frac{S_{\bar{k}i}}{S_{\bar{k}0}}
 & =
 s_{{R}_i^{(1)}}(q^{\hat{R}_k+\rho}) \, 
 s_{{R}_i^{(2)}}(q^{-\check{R}_k-\rho}) \, 
 \prod_{l,m=1}^N (q^{\hat{R}_{k,l}+\rho_l} + q^{-\check{R}_{k,m}-\rho_m})
 \nonumber \\
 & =
 s_{R_i} (q^{\hat{R}_k+\rho};q^{-\check{R}_k-\rho})
 \, .
\end{align}
Again this ratio is written in terms of the Schur polynomial itself, as
well as the bosonic $\U(N)$ theory.
Thus, applying the fusion formula \eqref{LR02}, we obtain
\begin{align}
 \frac{S_{\bar{k}i}}{S_{\bar{k}0}}
 \frac{S_{\bar{k}j}}{S_{\bar{k}0}}
 & =
 s_{R_i}(q^{\hat{R}_k+\rho};q^{-\check{R}_k-\rho}) \, 
 s_{R_j}(q^{\hat{R}_k+\rho};q^{-\check{R}_k-\rho}) \, 
 =
 \sum_\ell \mathpzc{N}_{~ij}^{~~\ell} \, 
 s_{R_\ell}(q^{\hat{R}_k+\rho};q^{-\check{R}_k-\rho}) 
 \, ,
 \label{V_formula3}
\end{align}
which shows the Verlinde formula for the $\U(N|M)$ theory.
This result suggests that we can construct a topological (knot)
invariant from the $\U(N|M)$ theory in a quite similar way to the bosonic
$\U(N)$ theory, at least for the maximal representations.


\section{Refinement of $\U(N|M)$ theory}\label{sec:refinement}

The result shown in Section~\ref{sec:op_formalism} suggests that the
$\U(N|M)$ theory follows most of the properties of the bosonic $\U(N)$
Chern--Simons theory.
In this Section, based on such a similarity, we try to apply
another interesting generalization of Chern--Simons theory to the $\U(N|M)$
theory. 

\subsection{Refined partition function}

\begin{figure}[t]
  \begin{center}
   \begin{tikzpicture}[node distance = 3cm, auto]
    
    \tikzstyle{block} = [thick, rectangle, draw, 
    text width=15em, text centered, rounded corners, minimum height=4em]

    \tikzstyle{block2} = [thick, rectangle, draw, 
    text width=16em, text centered, rounded corners, minimum height=21em]

    \node [block] (CS) {CS theory on $S^3$ \eqref{CS_pf}};
    \node [block, below of=CS] (CSlens) {CS theory on $S^3/\bZ_2$};
    \node [block, below of=CSlens] (ABJM) 
      {Supermatrix CS theory \eqref{ABJM_pf} \\ (ABJM theory)};

    \begin{scope}[node distance = 9cm]
     \node [block, right of=CS] (refCS) 
     {Ref. CS theory on $S^3$ \eqref{ref_CS_pf}};
    \end{scope}

    \node [block, below of=refCS] (refCSlens)
    {Ref. CS theory on $S^3/\bZ_2$ \eqref{ref_CS2_pf}};
    \node [block, below of=refCSlens] (refABJM) 
      {Ref.~supermatrix~CS~theory~\eqref{ref_ABJM_pf} \\ (Ref. ABJM theory)};

    \node [block2] (before) at (CSlens) {};
    \node [block2] (after) at (refCSlens) {};

    \draw [->,thick] (CS) -- (CSlens);
    \draw [->,thick] (CSlens) -- (ABJM);
    \draw [->,thick] (refCS) -- (refCSlens);
    \draw [->,thick] (refCSlens) -- (refABJM);
    \draw [->,ultra thick,blue] (before) -- (after);

    \node at (4.5,-2.5) {\textbf{Refinement}};
    \node at (.5,-1.5) [right] {$\bZ_2$ orbifold};
    \node at (.5,-4.5) [right] {Analytic cont.};
    \node at (9.5,-1.5) [right] {$\bZ_2$ orbifold};
    \node at (9.5,-4.5) [right] {Analytic cont.};

   \end{tikzpicture}
  \end{center}
  \caption{Chern--Simons theory to the refined supermatrix Chern--Simons
  (ABJM) theory. We apply the same way to obtain ABJM theory from
  Chern--Simons theory to the refined theory.}
  \label{CS_flow}
 \end{figure}
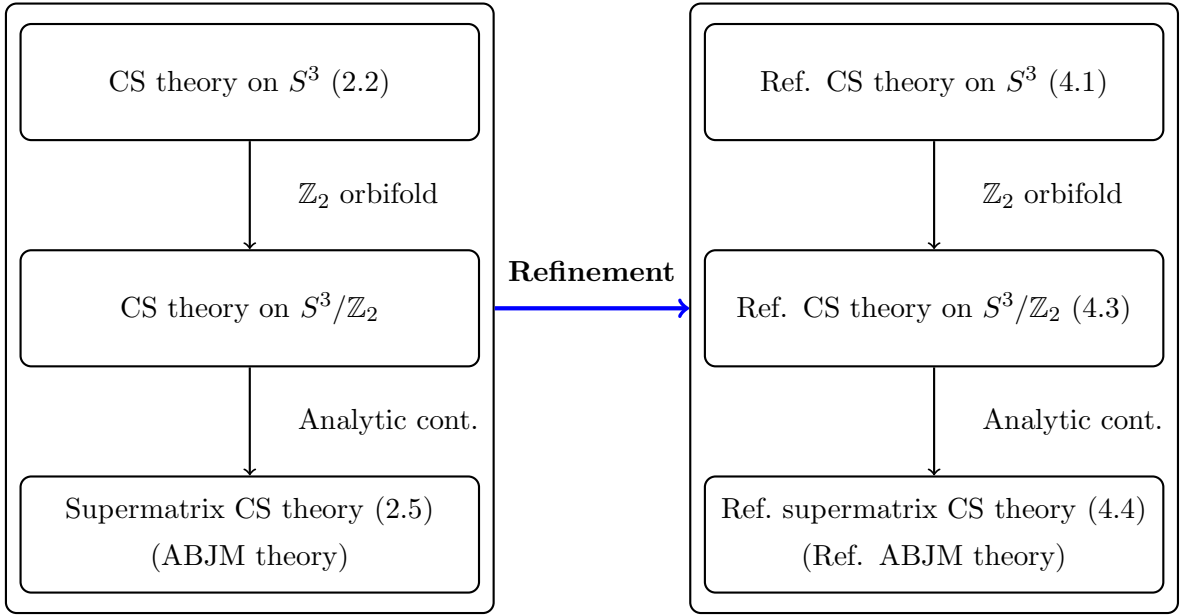

Based on the construction of Chern--Simons theory using topological
strings and its interpretation in M-theory, it has been proposed that the
refined index of M-theory defines the refined Chern--Simons
theory~\cite{Aganagic:2011sg}.
The most important example is the refined theory on the
three-sphere $S^3$.
However its construction can be applied to a wide range of 
three manifolds, by replacing the Calabi--Yau threefold and
the corresponding Lagrangian submanifold for M5-branes.
Indeed the ABJM theory is obtained from the Chern--Simons theory
on the lens space $S^3/\bZ_2$ through the analytic continuation.
In this sense, we can discuss the refined ABJM theory similarly using
the refined $\U(N|M)$ theory defined on $S^3/\bZ_2$ (see
Figure~\ref{CS_flow}).

As well as the bosonic $\U(N)$ Chern--Simons theory, the partition
function of the $\U(N)$ refined Chern--Simons theory on $S^3$ has a
(matrix) integral form,
\begin{align}
 \cZ_{\U(N)}^\text{Ref}(S^3;q,t)
 & =
 \frac{1}{N!} \int \prod_{i=1}^N \frac{\dif x_i}{2\pi} \,
 \nap^{-\frac{1}{2g_s} x_i^2} \,
 \prod_{i<j}^N
 \prod_{m=0}^{\beta-1}
 \left(
  \nap^{x_{ij}/2} q^{m/2} - \nap^{-x_{ij}/2} q^{-m/2}
 \right)^2
 \, ,
 \label{ref_CS_pf}
\end{align}
where we write $x_{ij}=x_i-x_j$, and the coupling constant is slightly
modified $q = \nap^{-g_s}$ with $g_s = 2\pi \im/(k+\beta N)$.
The remarkable modification is the measure part of this matrix integral,
which depends on the additional parameter $\beta$.
This expression corresponds to the situation such that $t=q^\beta$ with
$\beta \in \mathbb{N}$, and we focus basically on this case in the
following for simplicity.
For generic $\beta$, it is represented using infinite product,
\begin{align}
 &
 \prod_{i<j}^N
 \prod_{m=0}^\infty
 \frac{\nap^{x_{ij}/2} q^{m/2} - \nap^{-x_{ij}/2} q^{-m/2}}
      {\nap^{x_{ij}/2} t^{1/2}q^{m/2} - \nap^{-x_{ij}/2} t^{-1/2}q^{-m/2}}
 \nonumber \\
 & =
 \prod_{i<j}^N
 \prod_{m=0}^{\beta-1}
 \left(
  \nap^{x_{ij}/2} q^{m/2} - \nap^{-x_{ij}/2} q^{-m/2}
 \right)
 \quad \text{for} \quad t = q^\beta \quad (\beta \in \mathbb{N})
 \, .
\end{align}
It is easy to see that this is reduced to the non-refined Chern--Simons
partition function \eqref{CS_pf} by taking $\beta=1$, namely $t=q$.
As seen later, this modification corresponds to that for the Macdonald
polynomial from the Schur polynomial.

This refined Chern--Simons theory has essentially the same $\SL(2,\bZ)$
structure in its operator formalism as the non-refined Chern--Simons
theory~\cite{Aganagic:2011sg}.
This means that we can obtain the refined Chern--Simons theory on the
lens space $S^3/\bZ_2$ applying the same $\SL(2,\bZ)$ action to the
solid torus~\cite{Aganagic:2002wv}, which partly breaks the gauge symmetry.
Assuming that the system has the symmetry $\U(N+M)$ in the first place, which
shall be broken into $\U(N) \times \U(M)$, the partition function becomes
\begin{align}
 &
 \cZ_{\U(N)\times\U(M)}^\text{Ref}(S^3/\bZ_2;q,t)
 \nonumber \\
 & =
 \frac{1}{N!M!} \int 
 \prod_{i=1}^N \frac{\dif x_i}{2\pi} \, \nap^{-\frac{1}{2g_s} x_i^2} \,
 \prod_{j=1}^M \frac{\dif y_j}{2\pi} \, \nap^{-\frac{1}{2g_s} y_j^2} \,
 \prod_{m=0}^{\beta=1}
 \prod_{i=1}^N \prod_{j=1}^M
 \left(
  \nap^{\frac{1}{2}(x_i-y_j)} q^{m/2} + \nap^{-\frac{1}{2}(x_i-y_j)} q^{-m/2}
 \right)^2
 \nonumber \\
 & \hspace{5em} \times
 \prod_{m=0}^{\beta-1}
 \prod_{i<j}^N
 \left(
  \nap^{x_{ij}/2} q^{m/2} - \nap^{-x_{ij}/2} q^{-m/2}
 \right)^2
 \prod_{i<j}^M
 \left(
  \nap^{y_{ij}/2} q^{m/2} - \nap^{-y_{ij}/2} q^{-m/2}
 \right)^2
 \, .
 \label{ref_CS2_pf}
\end{align}
This is equivalent to the situation such that the second set of
variables is shifted due to the $\bZ_2$ discrete flat connection,
 $y_j \to y_j + \pi \im$.
Then, analytically continuating the gauge group rank $M \to - M$, which
is approved at least in a perturbative sense (see, for
example,~\cite{Dijkgraaf:2002pp}), and rescaling the coupling constant
$g_s = 2\pi \im /k$, we obtain the refined ABJM matrix model,
\begin{align}
 &
 \cZ_{\U(N|M)}^\text{Ref}(S^3;q,t)
 \nonumber \\
 & =
 \frac{1}{N!M!} \int 
 \prod_{i=1}^N \frac{\dif x_i}{2\pi} \, \nap^{-\frac{1}{2g_s} x_i^2} \,
 \prod_{j=1}^M \frac{\dif y_j}{2\pi} \, \nap^{\frac{1}{2g_s} y_j^2} \,
 \prod_{m=0}^{\beta=1}
 \prod_{i=1}^N \prod_{j=1}^M
 \left(
  \nap^{\frac{1}{2}(x_i-y_j)} q^{m/2} + \nap^{-\frac{1}{2}(x_i-y_j)} q^{-m/2}
 \right)^{-2}
 \nonumber \\
 & \hspace{5em} \times
 \prod_{m=0}^{\beta-1}
 \prod_{i<j}^N
 \left(
  \nap^{x_{ij}/2} q^{m/2} - \nap^{-x_{ij}/2} q^{-m/2}
 \right)^2
 \prod_{i<j}^M
 \left(
  \nap^{y_{ij}/2} q^{m/2} - \nap^{-y_{ij}/2} q^{-m/2}
 \right)^2
 \, .
 \label{ref_ABJM_pf}
\end{align}
For the simplest case $N=M=1$, we can exactly compute the
partition function and several expectation values with this model.
See Appendix~\ref{sec:Ref_Abelian} for details.
This is one-parameter generalization of the ABJM matrix model
\eqref{ABJM_pf}, and it is again reduced to the non-refined theory by taking
$\beta=1$.
We remark that the matrix measure of this partition function is not
written as a determinant anymore, due to the modification with the
additional parameter $t$, while the non-refined measure has the Cauchy
determinant expression.
This implies that the corresponding wavefunction for the refined theory
does not describe the free fermion system whose wavefunction has to be
expressed as a Slater determinant.

\subsection{Loop average}

For the refined $\U(N|M)$ theory \eqref{ref_ABJM_pf}, the most natural
observable is the Wilson loop operator as well as the non-refined
Chern--Simons theory.
However, in this case, the insertion of the loop gives rise to the
modified character of the corresponding group, which is expressed in
terms of the Macdonald polynomial~\cite{Aganagic:2011sg}.
Therefore, first of all, we have to introduce the $\U(N|M)$ version of
the Macdonald polynomial for our purpose.

The definition of the $\U(N|M)$ Macdonald polynomial should be given in
a similar way to the Schur polynomial, because it has to be reduced to
the Schur polynomial in the limit $t = q$.
The Macdonald polynomial is also a symmetric polynomial, which can
be expanded with the power-sum polynomials.
Thus, starting with the bosonic $\U(N+M)$ Macdonald polynomial, and
rewriting it in terms of the power-sum $\Tr U(x,y)^n$, one can obtain
the $\U(N|M)$ polynomial by replacing that power-sum with the $\U(N|M)$
counterpart, $\Str U(x;y)^n$.

Even with this definition, it is still difficult to write down the
explicit form of the polynomial for generic representations.
However, especially for the maximal representation, it is natural for
the $\U(N|M)$ Macdonald polynomial to have a similar factorization
formula as well as the Schur polynomial \eqref{Schur_fact02}.
It is because all the symmetric polynomials in the maximal
representation have to span the vector space, corresponding to the
subsector of the original $\U(N|M)$ representations, given by $\U(N)
\times \U(M) \subset \U(N|M)$.
This implies that it should be written as a product of the
$\U(N)$ and $\U(M)$ polynomials.
This speculation leads to the conjectural expression for the $\U(N|M)$
Macdonald polynomial in the maximal representation, which is a simple
generalization of the factorization formula \eqref{Schur_fact02}:
\begin{align}
 M_\lambda(x;y)
 & =
 M_{\lambda^{(1)}} (x) M_{\lambda^{(2)}} (y)
 \prod_{i=1}^N \prod_{j=1}^M \prod_{m=0}^{\beta-1}
 \left(
  x_i \, q^{m/2} + y_j \, q^{-m/2}
 \right)
 \, .
 \label{Mac_fact}
\end{align}
This is again the situation such that $t=q^\beta$ with
$\beta\in\mathbb{N}$, and its extension to arbitrary $\beta$ is
straightforward.

Starting with the conjectural formula \eqref{Mac_fact}, we can
compute several expectation values in the refined $\U(N|M)$
theory.
Let us assume $N=M$ in the following for simplicity.
The first example is the refined version of the inner product
\eqref{inner_prod}, which gives the Hermitian metric of the
corresponding Hilbert space \eqref{H_metric},
\begin{align}
 \langle R_i | R_j \rangle
 & =
 \frac{1}{N!^2} \int \prod_{i=1}^N 
 \frac{\dif x_i}{2\pi} \frac{\dif y_i}{2\pi} \,
 \prod_{m=0}^{\beta-1}
 \prod_{i<j}^N
 \left(
 \nap^{x_{ij}/2} q^{m/2} - \nap^{-x_{ij}/2} q^{-m/2}
 \right)^2
 \left(
 \nap^{y_{ij}/2} q^{m/2} - \nap^{-y_{ij}/2} q^{-m/2}
 \right)^2
 \nonumber \\
 & \hspace{.5em}
 \times
 \prod_{m=0}^{\beta-1}
 \prod_{i,j=1}^N
 \left(
  \nap^{\frac{1}{2}(x_i-y_j)} q^{m/2} + \nap^{-\frac{1}{2}(x_i-y_j)} q^{-m/2}
 \right)^{-2}
 M_{R_i}(\nap^{-x};\nap^{-y}) \,
 M_{R_j}(\nap^{x};\nap^{y}) 
 \, .
\end{align}
For the maximal representation, we can apply the factorization formula
\eqref{Mac_fact}, and thus this integral is factorized into two $\U(N)$
sectors.
Thus we can show the orthogonality 
\begin{align}
 g_{\bar{i}j}
 & = 
 g_{i^{(1)}} \, g_{i^{(2)}} \,
 \delta_{R_i^{(1)}R_j^{(2)}} \delta_{R_i^{(2)}R_j^{(2)}}
 \, ,
\end{align}
where the normalization factor becomes~\cite{Macdonald:1997}
\begin{align}
 g_{i}
 & =
 \prod_{m=0}^{\beta-1} \prod_{i<j}^N
 \frac{q^{\frac{1}{2}(R_i-R_j+m)}t^{\frac{1}{2}(j-i)}
       - q^{-\frac{1}{2}(R_i-R_j+m)}t^{-\frac{1}{2}(j-i)}}
      {q^{\frac{1}{2}(R_i-R_j-m)}t^{\frac{1}{2}(j-i)}
       - q^{-\frac{1}{2}(R_i-R_j-m)}t^{-\frac{1}{2}(j-i)}}
 \, .
\end{align}
It implies that, in contrast to the non-refined theory, it is not
orthonormal, but just orthogonal, as well as the refined $\U(N)$ theory,
and it is easy to see that it again becomes orthonormal $g_i = 1$ for
$\beta=1$.

We can similarly compute the two-point function especially with the
maximal representation, which gives the modular $S$-matrix element,
\begin{align}
 (TST)_{\bar{i}j}
 & =
 \frac{1}{N!^2} \int
 \prod_{i=1}^N \frac{\dif x_i}{2\pi} \frac{\dif y_i}{2\pi} \,
 \nap^{-\frac{1}{2g_s}(x_i^2-y_i^2)} \,
 \prod_{m=0}^{\beta-1}
 \prod_{i,j=1}^N
 \left(
  \nap^{\frac{1}{2}(x_i-y_j)} q^{m/2} + \nap^{-\frac{1}{2}(x_i-y_j)} q^{-m/2}
 \right)^{-2}
 \nonumber \\
 & \hspace{.5em}
 \times
 \prod_{m=0}^{\beta-1}
 \prod_{i<j}^N
 \left(
 \nap^{x_{ij}/2} q^{m/2} - \nap^{-x_{ij}/2} q^{-m/2}
 \right)^2
 \left(
 \nap^{y_{ij}/2} q^{m/2} - \nap^{-y_{ij}/2} q^{-m/2}
 \right)^2
 M_{R_i}(\nap^{-x};\nap^{-y}) \,
 M_{R_j}(\nap^{x};\nap^{y}) 
 \, .
\end{align}
In this case, due to the factorization of the Macdonald polynomial, we
can similarly apply the result for the bosonic refined $\U(N)$
theory~\cite{Etingof:1998ERA}, which yields
\begin{align}
 S_{\bar{i}j}
 & = 
 \frac{1}{k^N} \,
 M_{\hat{R}_i} (q^{\hat{R}_j} t^\rho) \,
 M_{\hat{R}_j} (t^\rho) \,
 M_{\check{R}_i} (q^{-\check{R}_j} t^{-\rho}) \,
 M_{\check{R}_j} (t^{-\rho})
 \nonumber \\
 & \hspace{5em}
 \times
 \prod_{m=0}^{\beta-1} \prod_{i<j}^N
 \left(
  t^{\rho_i} q^{m/2} - t^{\rho_j} q^{-m/2}
 \right)
 \left(
  t^{-\rho_i} q^{-m/2} - t^{-\rho_j} q^{m/2}
 \right)
 \, .
\end{align}
As well as the non-refined $\U(N|M)$ theory \eqref{TST_sSchur}, this
matrix element is completely factorized into the two $\U(N)$ sectors,
\begin{align}
 S_{\bar{i}j}^{\U(N|N)}(q,t)
 & =
 S_{\bar{\hat{i}}\hat{j}}^{\U(N)}(q,t) \times
 S_{\bar{\check{i}}\check{j}}^{\U(N)}(q^{-1},t^{-1})
 \, .
\end{align}

The expressions shown above can be obtained by a simple replacement of the $q$
parameter with $(q,t)$ in a proper way.
Furthermore, it is natural to expect that this refined $\U(N|M)$ theory also
involves the Verlinde formula at least for the maximal representations.
Thus this speculation leads to the following expression of the
$S$-matrix ratio in terms of the $\U(N|M)$ Macdonald polynomial,
\begin{align}
 \frac{S_{\bar{k}i}}{S_{\bar{k}0}}
 & =
 M_{R_i} (q^{\hat{R}_k}t^\rho;q^{-\check{R}_k}t^{-\rho})
 \, ,
\end{align}
which immediately yields the Verlinde formula
\begin{align}
 \frac{S_{\bar{k}i}}{S_{\bar{k}0}}
 \frac{S_{\bar{k}j}}{S_{\bar{k}0}}
 =
 \sum_\ell \mathpzc{N}_{~ij}^{~~\ell} \, 
 M_{R_\ell}(q^{\hat{R}_k}t^\rho;q^{-\check{R}_k}t^{-\rho})
 =
 \sum_\ell \mathpzc{N}_{~ij}^{~~\ell} \, 
 \frac{S_{\bar{k}\ell}}{S_{\bar{k}0}}
 \, ,
 \label{V_formula4}
\end{align}
where $\mathpzc{N}_{~ij}^{~~\ell}$ is the fusion coefficient for the
$\U(N|M)$ Macdonald polynomial.
In this case, this coefficient is not integer, but a rational function of
$q$ and $t$.
From this point of view, the refined $\U(N|M)$ theory also provides a
topological (knot) invariant in a three-manifold, which is the
categorified version of the non-refined $\U(N|M)$ invariant, given by the
Poincar\'e polynomial of the knot homology.

\subsection{Torus knot matrix model}

In addition to the partition function given in \eqref{ref_ABJM_pf}, we
can obtain another kind of integral by applying the $\SL(2,\bZ)$
transformation,
\begin{align}
 &
 \cZ_{\U(N|M)}^{\text{Ref},\, (P,Q)}(S^3;q,t)
 \nonumber \\
 & =
 \frac{1}{N!M!} \int 
 \prod_{i=1}^N \frac{\dif x_i}{2\pi} \, \nap^{-\frac{1}{2\hat{g}_s} x_i^2} \,
 \prod_{j=1}^M \frac{\dif y_j}{2\pi} \, \nap^{\frac{1}{2\hat{g}_s} y_j^2} \,
 \prod_{m=0}^{\beta=1}
 \prod_{i=1}^N \prod_{j=1}^M
 \left(
  \nap^{\frac{1}{2P}(x_i-y_j)} \hat{q}^{\frac{m}{2}} 
   + \nap^{-\frac{1}{2Q}(x_i-y_j)} \hat{q}^{-\frac{m}2}
 \right)^{-2}
 \nonumber \\
 & \hspace{6em} 
 \times
 \prod_{m=0}^{\beta-1}
 \prod_{i<j}^N
 \left(
  \nap^{\frac{1}{2P}x_{ij}} \hat{q}^{\frac{m}{2}} 
 - \nap^{-\frac{1}{2P}x_{ij}} \hat{q}^{-\frac{m}{2}}
 \right)^2
 \prod_{i<j}^M
 \left(
  \nap^{\frac{1}{2Q}y_{ij}} \hat{q}^{\frac{m}{2}} 
 - \nap^{-\frac{1}{2Q}y_{ij}} \hat{q}^{-\frac{m}{2}}
 \right)^2
 \, ,
 \label{PQ_ref_ABJM_pf}
\end{align}
where the coupling constant is rescaled, $\hat{q}=\nap^{-\hat{g}_s}$
with $\hat{g}_s = PQ g_s$.
Now one can compute the $(P,Q)$ torus knot average of the Wilson loop
operator with this integral, while the ordinary partition
function corresponding to $(P,Q)=(1,1)$ yields only the unknot average
of the loop operator.
The analysis of this integral would be interesting (and also complicated).
However we only show some results especially for $\U(1|1)$ theory in
Appendix~\ref{sec:Ref_Abelian}.

\section{Discussion}\label{sec:discussion}

In this paper we have studied the link average of the half-BPS Wilson
loop operator in ABJM theory based on the localization method.
The resultant expression is a simple generalization to the supermatrix
integral of the bosonic $\U(N)$ Chern--Simons theory.
We have computed the two- and three-link averages in the maximal
representation using the operator formalism inspired by the
three-dimensional topological field theory.
For the two-link average, which plays a role of the modular $S$-matrix,
we have obtained the factorization formula corresponding to
decomposition of the original supergroup into its subsectors, $\U(N)
\times \U(M) \subset \U(N|M)$.
For the three-link average, we have shown the Verlinde formula in the
$\U(N|M)$ theory.
We have also discussed a refinement of ABJM theory by
applying the argument based on topological string theory.
Applying the conjectural formula for the $\U(N|M)$ Macdonald polynomial,
we have derived the refined version of the link averages in several cases.

The operator formalism for the $\U(N|M)$ theory, as discussed in this paper,
has an analogous structure with the bosonic $\U(N)$ theory
for a class of the representation, which we call the maximal
representations.
This implies that we would have the chiral factorization on the boundary
of the three-manifold.
From the conformal field theoretical point of view, it is slightly an amazing
property, because such a factorization does not occur in a wide range
of CFTs involving internal supersymmetry. 
See, for example, a review article~\cite{Quella:2013oda} on this topic.
Therefore, in this sense, it would be interesting to see what happens in 
CFTs with internal supersymmetry with a sufficiently
large representation, e.g., the maximal representation discussed
in this paper.

The result obtained in this paper also suggests that we can similarly
construct the knot invariant associated with the supergroup $\U(N|M)$
even without using the matrix integral formula, namely just based on the
Skein relation.
At this moment it is not yet obvious whether it is possible for generic
representation.
However, for the maximal representation, we have almost the same property of
the link averages as the bosonic $\U(N)$ theory, and thus it suggests a
possibility to build a topological invariant with the $\U(N|M)$ theory.
Even in this case it is non-trivial, because originally ABJM
theory is just conformal, but not yet topological.
If we can successfully construct such an invariant, it would be
interesting to study the volume conjecture for hyperbolic knots,
since we can discuss the large representation limit with the maximal
representation.

For the refined $\U(N|M)$ theory, there are a lot of works to be
investigated.
First of all, it must be important to prove the conjectural
factorization formula for the $\U(N|M)$ Macdonald polynomial in the
maximal representation.
Since there is no determinant structure in this
case, its proof would be more difficult than the Schur polynomial, and
thus we should apply another approach, for example, based on the
differential (difference) operator acting on the symmetric polynomial.
A related issue is to explore an integrable model which is associated with
the $\U(N|M)$ Macdonald polynomial, namely the supergroup version of the
Ruijsenaars--Schneider model, and also its elliptic analog.
In addition, it might be possible to construct W-algebra based on this
kind of supergroup.

%


\subsubsection*{Acknowledgements}

We would like to thank B.~Eynard for the collaboration on the preceding
work, which leads to many ideas in this paper.
We are also grateful to V.~Pestun and S.~Ribault for useful discussions and
comments.
This work is supported in part by Grant-in-Aid for JSPS Fellows (\#25-4302).

\appendix
\section{$\U(N|M)$ Schur polynomial}\label{sec:Schur}

Let us summarize several properties of the $\U(N|M)$ Schur polynomial,
which are useful for the computation in the main part. 
For $\U(N|N)$ theory, there is a determinantal formula,
which is expressed in terms of the Frobenius
coordinate of the partition
$\lambda=(\alpha_1,\ldots,\alpha_{d(\lambda)}|\beta_1,\ldots,\beta_{d(\lambda)})$~\cite{Moens:2003JAC},
\begin{align}
 s_\lambda(u; v)
 & =
 \det_{1 \le i, j \le d(\lambda)}
 \left(
  \sum_{k,l=1}^N u_k^{\alpha_i} \left(C^{-1}\right)_{kl} v_l^{\beta_j}
 \right)
 \quad \text{with} \quad
 C_{ij} = \frac{1}{u_i+v_j}
 \, .
 \label{Schur_det1}
\end{align}
Here the matrix $C$ is called the Cauchy matrix, and $C^{-1}$ in the
formula is its inverse.
$d(\lambda)$ is the diagonal length of the partition $\lambda$, and
the Frobenius coordinates are associated with the partition in a relation of
$\alpha_i = \lambda_i - i$ and $\beta_i = \lambda^t_i-i$.
This formula means that this Schur polynomial is written as a determinant
of those in the hook representation
\begin{align}
 s_\lambda(u;v) & =
 \det_{1 \le i, j \le d(\lambda)}
 s_{(\alpha_i|\beta_j)} (u;v)
 \, ,
\end{align}
which is the Giambelli formula in this case.
From this formula, it is easily shown that $s_\lambda(u;v) = 0$ if
$d(\lambda) > N$, and let us call the representation corresponding to
the situation with $d(\lambda) = N$, which is equivalent to $\lambda_N
\ge N$, {\em a maximal representation} in this article.

\begin{figure}[t]
 \begin{center}
  \begin{tikzpicture}
   \path (0,0.35) node {$\lambda \ =$};
   \path (1.75,1.05) node {$N \times M$};
   \path (4.85,0.35) node {$\mu$};
   \path (2.75,-1.35) node {$\nu^t$};

   \draw 
   (1,0) -- (1,2.1) -- (2.5,2.1) -- (2.5,0) -- cycle ;

   \filldraw [fill=gray,fill opacity=.6,draw=black]
   (2.5,2.1) -- (5.2,2.1) -- (5.2,1.8) -- (4.3,1.8) -- 
   (4.3,1.2) -- (3.7,1.2) -- (3.7,0.9) -- (3.4,0.9) -- (3.4,0.6) --
   (2.8,0.6) -- (2.8,0.3) -- (2.5,0.3) -- cycle;

   \filldraw [draw=none,fill=gray,fill opacity=.6,draw=black]
   (1,0) -- (1,-1.5) -- (1.6,-1.5) -- (1.6,-0.9) -- (1.9,-0.9)
   -- (1.9,-0.6) -- (2.2,-0.6) -- (2.2,-0.3) -- (2.5,-0.3) 
   -- (2.5,0) -- cycle;

   \draw [<-,thick] (3.9,1.4) .. controls (4.1,0.7) and (4.2,0.7)
   .. (4.35,1) .. controls (4.5,1.2) and (4.6,1.2) .. (4.8,0.6);

   \draw [<-,thick] (1.75,-0.7) .. controls (2.4,-0.8) and (2.4,-0.9)
   .. (2.1,-1.1) .. controls (2,-1.2) and (2,-1.3) .. (2.5,-1.4);

  \end{tikzpicture}
 \end{center}
 \caption{The partition $\lambda = (14,11,11,9,8,6,5,5,4,3,2,2)$ 
 satisfying the maximal condition $\lambda_N \ge M$ for $N=7$ and
 $M=5$, which includes $\mu=(9,6,6,4,3,1,0)$ and $\nu^t=(5,4,3,2,2)$.
 }
 \label{fig:partition}
\end{figure}
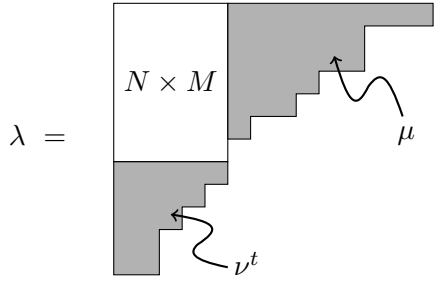

An important consequence of the determinantal formula \eqref{Schur_det1}
is that, in the maximal situation $d(\lambda)=N$, it is factorized
into the $\U(N)$ and $\U(M)$ Schur polynomials, 
\begin{align}
 s_\lambda(u;v) & = s_\mu(u) s_\nu(v) \prod_{i,j=1}^N (u_i + v_j)
 \, ,
 \label{Schur_fact01}
\end{align}
where the partitions in this formula are obtained from the original one,
\begin{align}
 \mu_i = \lambda_i - N \, , \qquad
 \nu_i = \lambda_i^t - N
 \qquad \text{for} \quad i = 1, \ldots, N \, ,
\end{align}
or equivalently,
\begin{align}
 \mu_i^t = \lambda_{i+N}^t \, , \qquad
 \nu_i^t = \lambda_{i+N} \, .
\end{align}
This kind of factorization can be also found in $\U(N|M)$ theory.
In this case, we call the situation with $\lambda_N \ge M$ a
maximal representation, and we have
\begin{align}
 s_\lambda(u;v) 
 & = s_\mu(u) s_\nu(v) \prod_{i=1}^N \prod_{j=1}^M (u_i + v_j)
 \, .
 \label{Schur_fact02}
\end{align}
Here the partitions are given by
$\mu_i = \lambda_i - M$ for $i=1,\ldots,N$ and
$\nu_i = \lambda_i^t - N$ for $i=1,\ldots,M$, or
$\mu_i^t=\lambda_{i+M}^t$ and $\nu_j^t=\lambda_{j+N}$.
We show an example of the maximal situation in Figure~\ref{fig:partition}.
This factorization reflects the fact that $\U(N|M)$ group contains
$\U(N)$ and $\U(M)$ groups as its subgroups, and the corresponding
characters are labeled by the partitions $\mu$ and $\nu$ obtained from
the original $\lambda$.

\section{Refined $\U(1|1)$ theory}\label{sec:Ref_Abelian}

In this Appendix we show several explicit results for the simplest
situation in the refined $\U(N|M)$ theory with $N = M = 1$.
Let us first compute the partition function for the $(P,Q)$ torus knot
\eqref{PQ_ref_ABJM_pf}, which includes the original situation
\eqref{ref_ABJM_pf} as $(P,Q) = (1,1)$,
\begin{align}
 \cZ_{\U(1|1)}^{\text{Ref},\,(P,Q)} 
 & =
 \int \frac{\dif x}{2\pi} \frac{\dif y}{2\pi} \,
 \nap^{-\frac{1}{2\hat{g}_s}(x^2-y^2)} \,
 \prod_{m=0}^{\beta-1}
 \left(
  2 \cosh \frac{x-y-m\hat{g}_s}{2P}
 \right)^{-1}
 \left(
  2 \cosh \frac{x-y-m\hat{g}_s}{2Q}
 \right)^{-1}
 \, .
\end{align}
Applying the Fourier transformation formula
\begin{align}
 \frac{1}{2 \cosh w}
 & =
 \int \frac{\dif z}{2\pi} \,
 \frac{\nap^{2 \im w z / \pi}}{\cosh z}
 \, ,
\end{align}
we have an expression,
\begin{align}
 \int \frac{\dif x}{2\pi} \frac{\dif y}{2\pi} 
 \prod_{m=0}^{\beta-1}
 \frac{\dif z_m}{2\pi} \frac{\dif w_m}{2\pi} \,
 \nap^{-\frac{1}{2\hat{g}_s}(x^2-y^2)} \,
 \prod_{m=0}^{\beta-1}
 \frac{\nap^{\frac{\im}{\pi}(x-y-m \hat{g}_s)(z_m/P+w_m/Q)}}
      {\cosh z_m \, \cosh w_m}
 \, .
\end{align}
Integrating out $x$ and $y$ variables first, we obtain
\begin{align}
 \cZ_{\U(1|1)}^{\text{Ref},\,(P,Q)} 
 & =
 \frac{PQ}{k}
 \prod_{m=0}^{\beta-1}
 \left(
  q^{Pm/2} + q^{-Pm/2}
 \right)^{-1}
 \left(
  q^{Qm/2} + q^{-Qm/2}
 \right)^{-1}
 \, .
\end{align}
By taking $(P,Q)=(1,1)$ and the non-refined limit $\beta=1$, it
reproduces the known result for the non-refined $\U(1|1)$ theory,
\begin{align}
 \cZ_{\U(1|1)}^{\text{Ref},\,(P,Q)} 
 \ \longrightarrow \
 \cZ_{\U(1|1)}^{(1,1)} 
 = \frac{1}{4k}
 \, .
\end{align}

Let us then compute the unknot average with $(P,Q) = (1,1)$.
In this case, only the hook representation is possible for the $\U(1|1)$
Macdonald polynomial as well as the Schur polynomial (otherwise it
trivially vanishes), and thus the conjectural formula \eqref{Mac_fact} yields
\begin{align}
 M_{(a|b)}(\nap^x;\nap^y)
 & =
 \nap^{ax+by}
 \prod_{m=0}^{\beta-1}
 \left(
  \nap^{x_i} q^{m/2} + \nap^{y_j} q^{-m/2}
 \right)
 \nonumber \\
 & =
 \nap^{(a+\frac{\beta}{2})x + (b+\frac{\beta}{2})y}
 \prod_{m=0}^{\beta-1}
 2 \cosh \left( \frac{x-y-mg_s}{2} \right)
 \, .
\end{align}
Thus the average of this Macdonald polynomial is given by
\begin{align}
 \Big< M_{(a|b)} (\nap^x;\nap^y) \Big>
 &
 =
 \frac{1}{\cZ_{\U(1|1)}^{\text{Ref},\, (1,1)}}
 \int \frac{\dif x}{2\pi} \frac{\dif y}{2\pi} \,
 \nap^{-\frac{1}{2\epsilon}(x^2-y^2)} \,
 \nap^{(a+\frac{\beta}{2})x+(b+\frac{\beta}{2})y} \,
 \prod_{m=0}^{\beta-1}
 \left( 2 \cosh \left( \frac{x-y-m \epsilon}{2} \right) \right)^{-1}
 \nonumber \\
 & =
 q^{\frac{1}{2}(a+b+\beta)(a-b)} \,
 \prod_{m=0}^{\beta-1}
 \frac{\left(q^{-m/2} + q^{m/2}\right)^2}
      {q^{\frac{1}{2}(a+b+\beta-m)}
     + q^{-\frac{1}{2}(a+b+\beta-m)}}
 \, .
\end{align}
This expectation value should give the unknot invariant, especially the
Poincar\'e polynomial of the corresponding knot homology,
and the first factor can be interpreted as the framing factor.

We can similarly compute the $(P,Q)$ torus knot average with respect to
the partition function \eqref{PQ_ref_ABJM_pf}.
Now we rewrite the $\U(1|1)$ Macdonald polynomial using the $q$-binomial
formula,
\begin{align}
 M_{(a|b)}(\nap^x;\nap^y)
 & =
 \sum_{\ell=0}^\beta
 q^{\frac{1}{4}\beta(\beta-\ell) - \frac{1}{2}\ell(\ell-1)}   
 \binom{\beta}{\ell}_{q} \,
 \nap^{(a + \beta - \ell)x + (b + \ell)y}
 \, ,
\end{align}
where the $q$-binomial coefficient is given by
\begin{align}
 \binom{\beta}{\ell}_{q} 
 & =
 \frac{(1-q^\beta)(1-q^{\beta-1})\cdots(1-q^{\beta-\ell+1})}
      {(1-q)(1-q^2)\cdots(1-q^{\ell})}
 \nonumber \\
 & =
 \frac{(q^{\ell+1};q)_\infty (q^{\beta-\ell+1};q)_\infty}
      {(q^{\beta+1};q)_\infty (q;q)_\infty}
 \, .
\end{align}
Thus, if we can compute the expectation value of the monomial $\nap^{(a
+ \beta - \ell)x + (b + \ell)y}$ with respect to the torus knot matrix
model, we obtain the torus knot average.
Now this average is given by
\begin{align}
 &
 \Big<  \nap^{(a + \beta - \ell)x + (b + \ell)y} \Big>_{(P,Q)}
 \nonumber \\
 & =
 {q}^{-\frac{PQ}{2}(a+\beta-\ell)^2-\frac{1}{2}(b+\ell)^2} \,
 \prod_{m=0}^{\beta-1}
 \frac{q^{Pm/2} + q^{-Pm/2}}
      {{q}^{\frac{P}{2}(a+b+\beta-m)} + {q}^{-\frac{P}{2}(a+b+\beta-m)}}
 \frac{q^{Qm/2} + q^{-Qm/2}}
      {{q}^{\frac{Q}{2}(a+b+\beta-m)} + {q}^{-\frac{Q}{2}(a+b+\beta-m)}}
 \, .
\end{align}
Note that the $\ell$-dependence is only found in the framing factor
in this expression, and thus we obtain the $(P,Q)$ torus knot average of
the $\U(1|1)$ Macdonald polynomial,
\begin{align}
 \Big< M_{(a|b)}(\nap^x;\nap^y) \Big>_{(P,Q)}
 & =
 \prod_{m=0}^{\beta-1}
 \frac{q^{Pm/2} + q^{-Pm/2}}
      {{q}^{\frac{P}{2}(a+b+\beta-m)} + {q}^{-\frac{P}{2}(a+b+\beta-m)}}
 \frac{q^{Qm/2} + q^{-Qm/2}}
      {{q}^{\frac{Q}{2}(a+b+\beta-m)} + {q}^{-\frac{Q}{2}(a+b+\beta-m)}}
 \nonumber \\
 & \hspace{5em} \times
 \sum_{\ell=0}^\beta
 q^{\frac{1}{4}\beta(\beta-\ell) - \frac{1}{2}\ell(\ell-1)}   
 \binom{\beta}{\ell}_{q} \,
 {q}^{-\frac{PQ}{2}(a+\beta-\ell)^2-\frac{1}{2}(b+\ell)^2} 
 \, .
\end{align}
We remark that this expression is symmetric under $P \leftrightarrow Q$
and $q \leftrightarrow q^{-1}$ apart from the framing factor.
Furthermore, from the view point of the Macdonald polynomial, it is
natural to have a symmetry of $q \leftrightarrow t$, which is
not obvious in this formula.
It would be convenient to rewrite it using infinite product to see such
a duality.


\bibliographystyle{ytphys}
\bibliography{/Users/k_tar/Dropbox/etc/conf}

\end{document}